\documentclass[%
 reprint,
 amsmath,amssymb,
 aps, prl
]{revtex4-1}

\usepackage{graphicx}
\usepackage{dcolumn}
\usepackage{bm}

\begin{document}
	
	\preprint{APS/123-QED}
	
	\title{Measuring anomalous heating in a planar ion trap with variable ion-surface separation}
	
	\author{Ivan A. Boldin}
	\email{ivan.boldin@uni-siegen.de}
	\author{Alexander Kraft}%
	\author{Christof Wunderlich}
	\email{christof.wunderlich@uni-siegen.de}
	\affiliation{%
		Department Physik, Naturwissensch\"aftlich-Technische Fakult\"at, Universit\"at Siegen, 57068 Siegen, Germany\\
	}%
	
	\date{\today}
	
	\begin{abstract}
		Cold ions trapped in the vicinity of conductive surfaces experience heating of their oscillatory motion. Typically, the rate of this heating is orders of magnitude larger than expected from electric field fluctuations due to thermal motion of electrons in the conductors. This effect, known as anomalous heating, is not fully understood. One of the open questions is the heating rate's dependence on the ion-electrode separation. We present a direct measurement of this dependence in an ion trap of simple planar geometry. The heating rates are determined by taking images of a single $^{172}$Yb${^+}$ ion's resonance fluorescence after a variable heating time and deducing the trapped ion's temperature from measuring its average oscillation amplitude. Assuming a power law for the heating rate vs. ion-surface separation dependence, an exponent of -3.79 $\pm$ 0.12 is  measured. 
	\end{abstract}
	
	\maketitle
	
	
	Electric field noise in close proximity to metal surfaces is an important issue in various fields of experimental physics, such as measuring weak forces in scanning probe microscopy \cite{dorofeyev1999brownian, stipe2001noncontact} or for Casimir effect studies \cite{speake2003forces, kim2010surface}, gravitational wave detection \cite{pollack2008temporal}, and experiments on the gravitational properties of charged particles \cite{darling1992fall}. In experiments with cold trapped ions such noise results in excitation (also termed heating) of the ions' motional degrees of freedom \cite{brownnutt2015ion}. In realizations of quantum information processing based on trapped ions, this heating can become a major source of decoherence \cite{brownnutt2015ion, schneider1999decoherence, leibfried2003quantum}. 
	
	Experiments have shown that the observed heating rate is orders of magnitude greater than would be caused by thermal motion of electrons in the conductors (i.e. Johnson noise) \cite{turchette2000heating, deslauriers2006scaling}. This high heating rate is mostly associated with surface contamination and surface imperfections, as surface treatment is known to be able to reduce the heating rate significantly \cite{hite2012100, daniilidis2014surface}. However, its mechanism is not fully understood, and thus this effect is referred to as anomalous heating; a recent review of experimental and theoretical studies of this phenomenon is given in \cite{brownnutt2015ion}. A comparison of experiments, employing different types and sizes of ion traps, shows that the anomalous heating rate grows fast as the ion-electrode separation decreases \cite{brownnutt2015ion}. Therefore, anomalous heating is particularly prominent for microfabricated planar ion traps \cite{chiaverini2005}, where this separation can be as small as tens of micrometers. Microfabricated traps are central for the realization of scalable quantum information processing with trapped ions \cite{chiaverini2005,seidelin2006microfabricated, wang2010demonstration, hughes2011microfabricated, ospelkaus2011microwave, wilpers2012monolithic, kunert2014, wilson2014tunable, monroe2014large, brandl2016cryogenic, herold2016universal}, and, therefore, in addition to its fundamental interest, it is of particular importance to characterize and understand anomalous heating.
	
	Though electric field noise-induced heating of ion motion has been studied in many experimental and theoretical works over the last years \cite{brownnutt2015ion}, one of the still open questions regarding anomalous heating is its dependence on the ion-electrode separation. In addition to being of practical use for ion trap design, knowing this dependence can confirm or contradict various existing theoretical models of anomalous heating. Usually, a single ion trap does not offer a possibility to vary the ion-electrode separation. A possible way to measure the heating rate dependence on ion-electrode distance is to compare ion heating rates in different traps with different electrode geometries; this was done for two traps that were scaled versions of each other \cite{turchette2000heating}. However, ion heating rates often show poor reproducibility even between identically designed traps, and therefore such experiments may not be ideal to probe this dependence.
	
	To our knowledge there have been two direct measurements of heating rate dependence on the distance between trapped ion and the nearest electrode \cite{deslauriers2006scaling, hite2017measurements}. One study \cite{deslauriers2006scaling} was carried out in a Paul trap with an ion trapped between two needle-shaped electrodes with the distance between them varying from 38 to 220 $\mu$m. Fitting the heating rate vs. distance to the needles with a power law gave an exponent of -3.5$\pm$0.1. Another experiment \cite{hite2017measurements} was done in a ''stylus trap'' \cite{arrington2013micro} with a flat electrode placed opposite the trap at a variable distance to it. The authors concluded that the flat electrode does not give a significant contribution to the heating rate, and for the dependence of the heating rate on the distance between an ion and the stylus trap, a power law with an exponent of -3.1 was obtained.
	
	In both experiments, modeling the dependence of the heating rate on the distance to the electrodes is likely to require geometrical factors. For a theoretical interpretation of this dependence, it would be beneficial to exclude the factor of trap geometry in this dependence by performing the measurements in a trap of simple geometry such as a planar electrode trap. In this work we measure the heating rates in a single micro-structured planar electrode ion trap with the ability to vary the ion-surface separation. As all electrodes of the trap lie in one plane, and the gaps between them are much smaller than the ion-surface distance, the trap can be viewed as an infinite plane when considering possible theoretical models for anomalous heating. The electrode configuration of planar traps is also of particular practical importance as such traps are widely used in experiments on quantum information processing and in other experiments.
	
	We trap single $^{172}$Yb${^+}$ ions in a 5-electrode-type surface trap \cite{chiaverini2005} made of gold electroplated on a sapphire substrate \cite{kunert2014}. The variation of the trapping height is achieved via applying a radio frequency (RF) voltage of variable amplitude to the central electrode of the trap, in addition to the main RF drive of the trap (Fig.~\ref{fig:png1}). This method of controlling the trapping height was also implemented in a larger-scale surface trap \cite{cetina2007bright} and in a circular electrode point-trap \cite{kim2010surface2}. A sinusoidal RF drive of 13.5 MHz is supplied to a helical resonator \cite{kunert2014}, and after the resonator a custom-made capacitive voltage divider splits the RF signal into two signals of the same phase.  By tuning the capacitances of the divider, the ratio of their amplitudes can be varied. This setup allows for varying the ion-surface distance, $h$, in the range of approximately 45 to 155 $\mu$m. 
	
	\begin{figure}[b]
		\includegraphics[width=5cm]{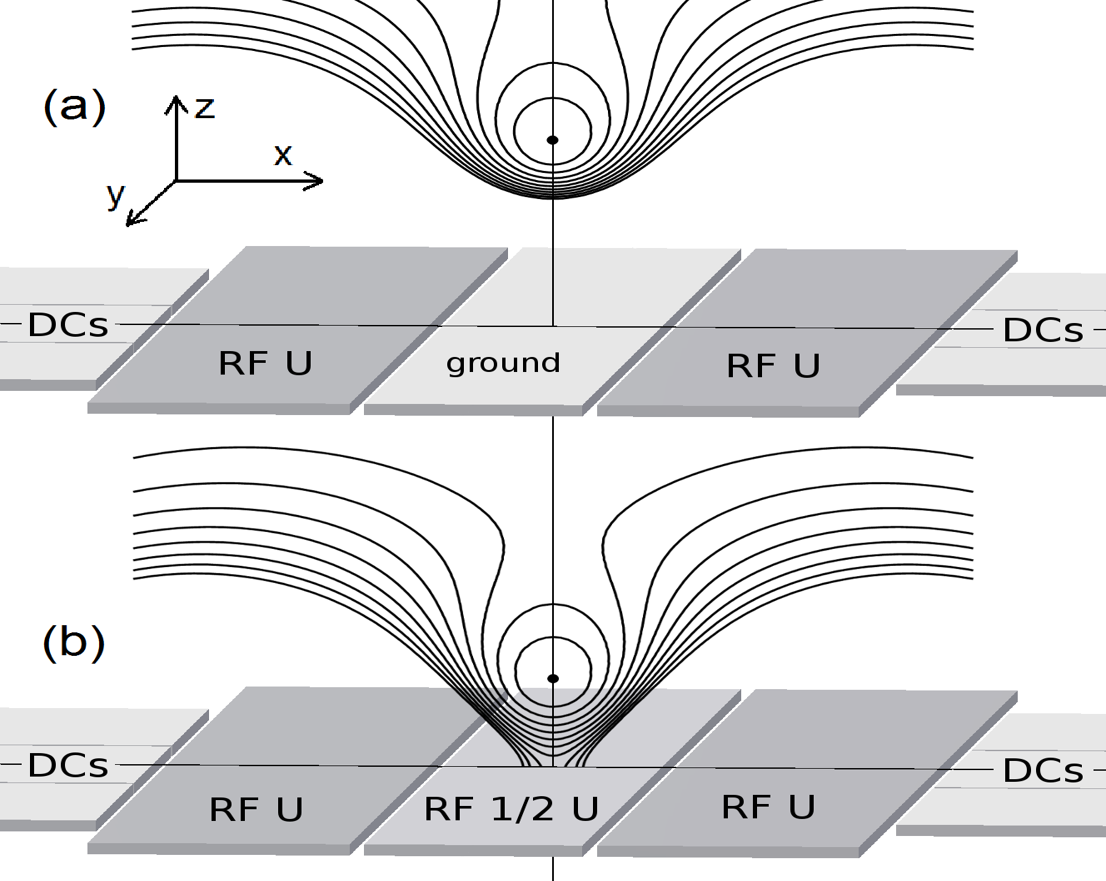}
		\caption{\label{fig:png1} Trapping height variation principle. a) schematic representation of our surface trap in maximum trapping height mode. An RF voltage of amplitude $U$ is applied to the electrodes shown in dark grey, the central electrode is grounded. Equipotential lines of the effective potential in the plane perpendicular to the trap axis (y-direction) are shown. b) the trap and effective potential when an additional RF voltage of amplitude $1/2\ U$ is applied to the central electrode to reduce the trapping height.}
	\end{figure}
	
	A single trapped ion, Doppler-cooled on the S$_{1/2}$ -- P$_{1/2}$ resonance near 369.5 nm, is used as a probe.  Resonance fluorescence at this wavelength is collected using a custom-made objective with numerical aperture 0.4 \cite{schneider2007} allowing for near diffraction limited spatial resolution, and is registered employing an EMCCD camera. The trap is kept in vacuum better then $3 \cdot 10^{-11}$ Torr. We measure heating of the axial mode of secular ion motion, that is, along the y-axis shown in Fig.~\ref{fig:png1}a. 
	
	In order to measure the motional heating rates, first we implement the so-called recooling method \cite{wesenberg2007fluorescence, brama2012heating}. In this method an ion is Doppler-cooled to a temperature of order mK, then the cooling laser is blocked to let the ion heat up; finally, after heating time $T_H$ (on the order of seconds), Doppler-cooling is switched back on and the ion's photon scattering rate is observed with high temporal resolution while the ion is being cooled back to its initial equilibrium temperature. The time evolution of the fluorescence rate as a function of time can be theoretically modeled, and the average energy of an ion after heating is obtained by fitting the model to the experimental data. It was previously shown that the recooling method gives results that are in agreement with the sideband method \cite{epstein2007simplified}, even though $T_H$ differs by a few orders of magnitude between these two methods.
	
	In addition to employing the recooling method to determine the ion's average kinetic energy after heating, we determine this energy by measuring the average ion oscillation amplitude after heating (this method is described in the next paragraph). From a comparison of these two methods we conclude that the recooling method in this experiment overestimates the energy in the axial mode after heating  by about an order of magnitude. This can be explained as follows: An important feature of the recooling method is that it assumes that only the axial mode of motion is excited \cite{wesenberg2007fluorescence}. However, if external noise (such as technical noise or Johnson noise, see Supplemental Material \cite{sm}) is present on the electrodes of the trap, then the axial and radial components of the fluctuating electric field due to this noise can differ significantly, and there is no a priori reason to assume that after a heating time the radial oscillation energies will be much lower than the axial one. This external noise is not the dominant source of ion heating in the axial direction in our experiment; we conclude this from comparing the measured heating rate as a function of ion-surface distance with the expected heating rate dependence on this distance in the case of external noise causing this heating (See Supplemental Material \cite{sm}). 
	
	Therefore, we use a different method of measuring an ion's motional energy after heating. The same experimental sequence as in the recooling method is carried out, but instead of the fluorescence rate's dependence on time, one measures the spread of the ion image in the axial direction after the heating time, $T_H$. The spatial extension of the resonance fluorescence was used for thermometry of a single ion in thermal equilibrium \cite{knunz2012sub}. Here, we record time-resolved ion images. The EMCCD camera is directed along the z-axis and obtains images in the x-y plane (Fig.~\ref{fig:png1}).  This experimental sequence is repeated typically around 200 times and the camera images for each time frame are summed. An example of the recorded ion images summed over the x-direction for different delays after switching on the cooling laser is shown in Fig.~\ref{fig:png2}. We assume that, after being heated, a harmonically trapped ion has a thermal energy distribution with average energy $E$ that we want to measure. Then it has the following probability distribution, $\rho$, in phase space:
$\rho=\frac{\omega}{2 \pi E} \exp\left(-\frac{m \omega^2 x^2 + p^2/m}{2 E}\right)$,
where $m$ is the ion's mass, $\omega$ is its axial harmonic oscillator frequency, and $x$ and $p$ are position and momentum along the axial direction, respectively.
	The ion's fluorescence rate, $F$, depends on $p$ because of the Doppler shift, and can be approximated as not depending on $x$, if the oscillation amplitude is much smaller than the spatial width of the laser beam exciting resonance fluorescence. Then, the distribution of resonance fluorescence over $x$, $\rho_F(x)$, is given by
$\rho_F(x) = \frac{\omega}{2 \pi E} \int\limits_{-\infty}^{\infty} \exp\left(-\frac{m \omega^2 x^2 + p^2/m}{2 E}\right) F(p) dp \\ = \exp\left(-\frac{m \omega^2 x^2}{2 E}\right) \cdot \text{const} = \exp\left(-\frac{x^2}{2 \sigma^2}\right) \cdot \text{const} $
that is, it is a Gaussian distribution with the root mean square (RMS) width $\sigma$ and $\sigma^2 = E / m \omega^2$.
	
	\begin{figure}[b]
		\includegraphics[width=\columnwidth]{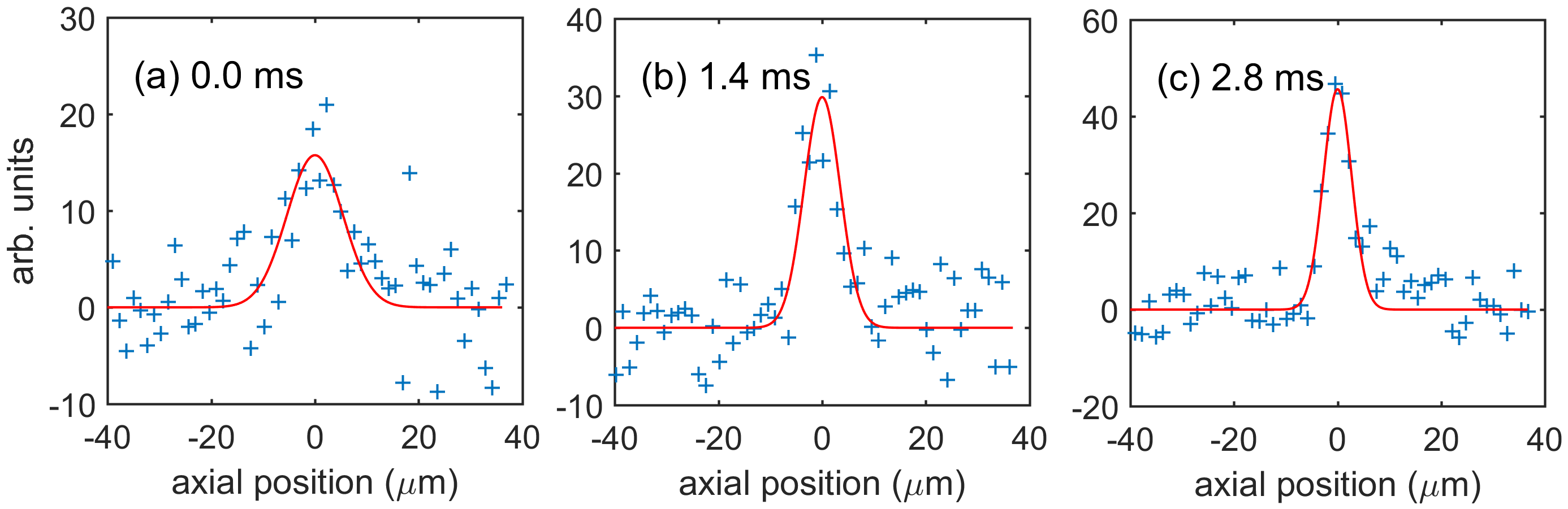}
		\caption{\label{fig:png2} Ion images summed over the radial direction (blue crosses) and Gaussian fits (red solid lines) at $t=0, 1.4$, and $2.8$ ms after opening the cooling laser. The heating time $T_H$ is 15 s, exposure time is 0.2 ms, 209 experimental runs are summed.}
	\end{figure}
	
	\begin{figure}[b]
		\includegraphics[width=6cm]{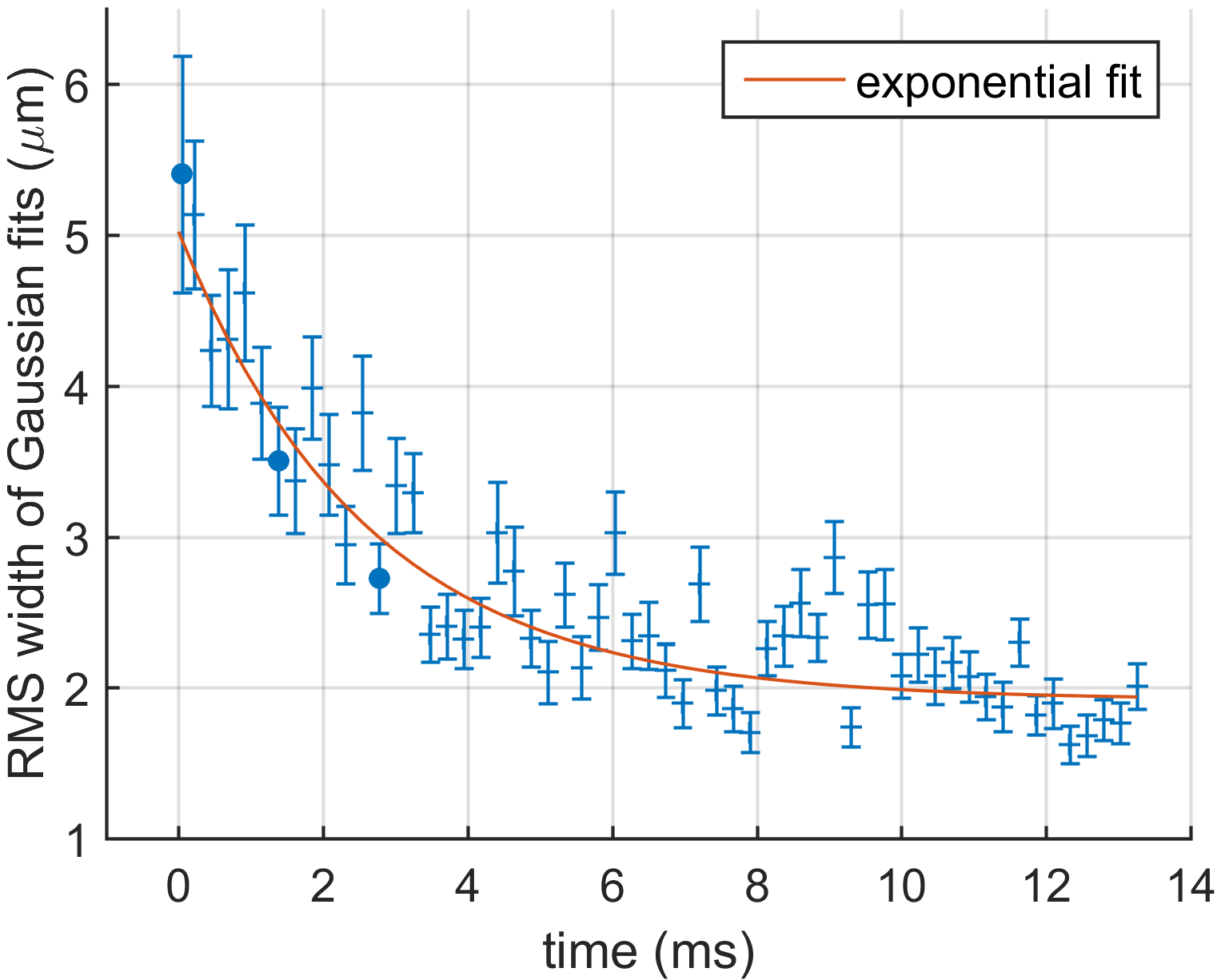}
		\caption{\label{fig:png3} Spread of the resonance fluorescence of an ion in axial direction expressed as RMS width $\sigma$ of the fitted Gaussians (some of which are shown in Fig.~\ref{fig:png2}, the corresponding points are marked with filled circles) depending on time. An exponential fit is shown in red.}
	\end{figure}
	
	In principle, in order to determine $E$, one could just take an image of the ion's resonance fluorescence right at the moment when the cooling laser is switched back on ($t$=0), sum it over the radial direction and average over many experimental runs (Fig.~\ref{fig:png2} a)). Then a Gaussian fit of this snapshot of the spatial distribution of the ion's resonance fluorescence along the axial direction could be done, and thus $\sigma$ and $E$ be extracted. Here, in order to improve the precision in determining $\sigma$, we measure the dependence of the RMS widths of the Gaussian fits on time {\it after} the cooling laser has been switched back on  (Fig.~\ref{fig:png2} b) for $t=1.4$ ms; Fig.~\ref{fig:png2} c) for $t=2.8$ ms) and extrapolate it to zero time to obtain $\sigma$ and $E$ (Fig.~\ref{fig:png3}, for 0 ms $\leq t \leq $ 13.3 ms) using an exponential fit of the data in Fig.~\ref{fig:png3}.
	
	The method based on the analysis of ion images can be advantageous as compared with simple recooling for measuring heating rates because: a) no assumptions about the radial motion are needed as the axial amplitude is measured directly; b) The fluorescence rate vs. time strongly depends on laser power and laser detuning, so for the simple recooling method these parameters should be kept constant with high precision, which can be experimentally challenging, while for measuring axial amplitudes directly this precision is not needed.
	
	Using the method described above, we measure the average energy $E$ as a function of $T_H$ with all other parameters held fixed. The ion heating rate for each height $h$ and frequency $\omega$ is obtained by three measurements of average energy for each of three heating times $T_H$. Then a linear fit of $E$ vs. $T_H$ yields the heating rate for a given parameter set. This, in turn, allows for determining the dependence of the heating rate on $h$ and $\omega$.
	
	The heating rates are measured for the range of trapping heights $h$ from 61$\pm$1.5 $\mu$m to 154$\pm$1.5 $\mu$m. The axial secular frequency $\omega$ is kept at $2\pi\times196\pm$2 kHz for various trapping heights and is varied from $2\pi\times90\pm$0.9 kHz to $2\pi\times290\pm$3 kHz for the frequency dependence measurement by adjusting the voltages applied to the DC electrodes of the trap. The radial frequencies are in the range from 1.0 to 1.4 MHz for all data points. Micromotion minimization was carried out before every measurement using the method of observing an ion’s positions while varying the amplitude of the RF drive \cite{gloger2015ion}. The cooling laser power is chosen so that the saturation parameter for the 369.5 nm cooling transition is in the range of 1 -- 1.5. The line width of this transition is 19.8 MHz and the detuning was chosen between 5 and 10 MHz.
	
	A linear  dependence of $E$ on $T_H$ is assumed here, because as long as $E$ (on the order of 1 K in these experiments) is much lower than the temperature of the reservoir (the trap that is kept at room temperature), the heating rate is constant in time \cite{brownnutt2015ion}. The heating times $T_H$ were chosen such that ions acquire approximately the same average oscillation amplitudes for all trapping heights $h$ and frequencies $\omega$. Because of this, if a systematic error on the heating rate would be present, its effect on the measured heating rate vs. $h$ or $\omega$ would be reduced. $\sigma$  was typically in the range from 4 to 8 $\mu$m, and $T_H$ ranged from 1 s for the highest heating rates to 90 seconds for the lowest. The heating rates, $P$, are presented in Kelvin per second,  not to be confused with $\Gamma$ in quanta per second -- they are related as $P=\Gamma \hbar \omega / k_B$, where $k_B$ is Boltzmann's constant. $P$ is also related to the spectral density of the electric field noise $S_E$ as $P = \frac{e^2}{4 m  k_B} S_E$ \cite{turchette2000heating}, where $e$ is the ion's net charge. The lowest heating rate, 0.0287$\pm$0.0032 K/s, measured at $\omega=2\pi\times290\pm$3 kHz and $h=134\pm$1.5 $\mu$m, corresponds to $\Gamma$=3.1$\pm$0.35 quanta/ms and  $S_E = 1.78 \pm 0.20 \cdot 10^{-11} \ \textrm{V}^2 \textrm{m}^{-2} \textrm{Hz}^{-1}$.
	
	The heating rate $P$ as a function of $\omega$ and  $h$ is shown in Fig.~\ref{fig:png4} and Fig.~\ref{fig:png5}, respectively. The results are well fitted by power functions, and by doing so one obtains the power laws for both dependencies: $P \propto h^{-3.79 \pm 0.12}$ and $P \propto \omega^{-1.13 \pm 0.10}$. The major contribution to the error (one standard deviation) in the exponent is due to the uncertainty in determining the Gaussian width in Fig.~\ref{fig:png2}. It accounts for from 6.3\% to 10.9\% (8.4\% on average) error in the heating rate. A 1\% error in axial frequency measurement yields 2\% error in the heating rate. The trapping height $h$ is determined with $\pm$1.5 $\mu$m precision.
	
	\begin{figure}[b]
		\includegraphics[width=7cm]{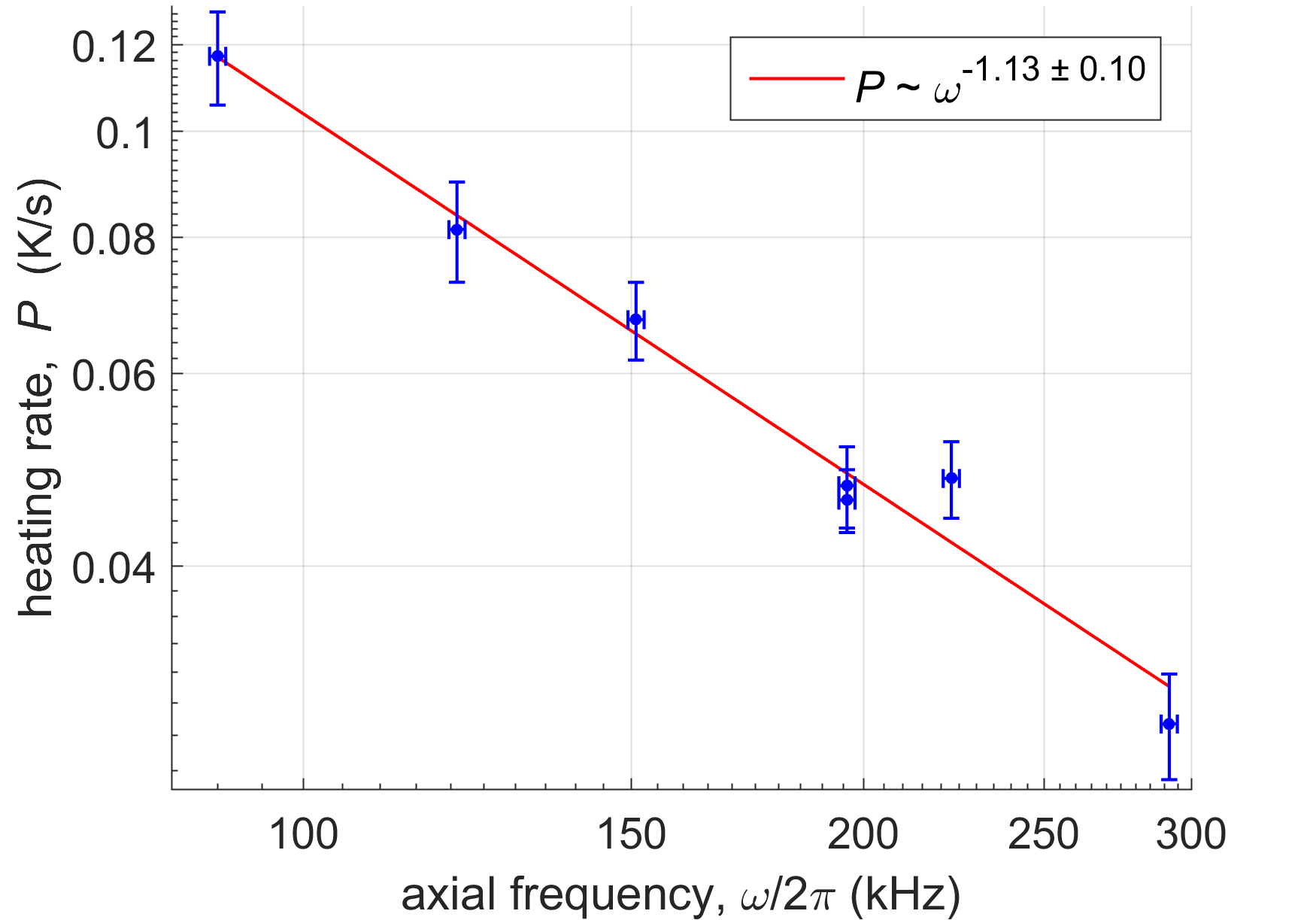}
		\caption{\label{fig:png4} Heating rate (in Kelvin per second) dependence on the axial secular frequency. The trapping height is 134$\pm$1.5 $\mu$m for all data points.}
		
		\includegraphics[width=7cm]{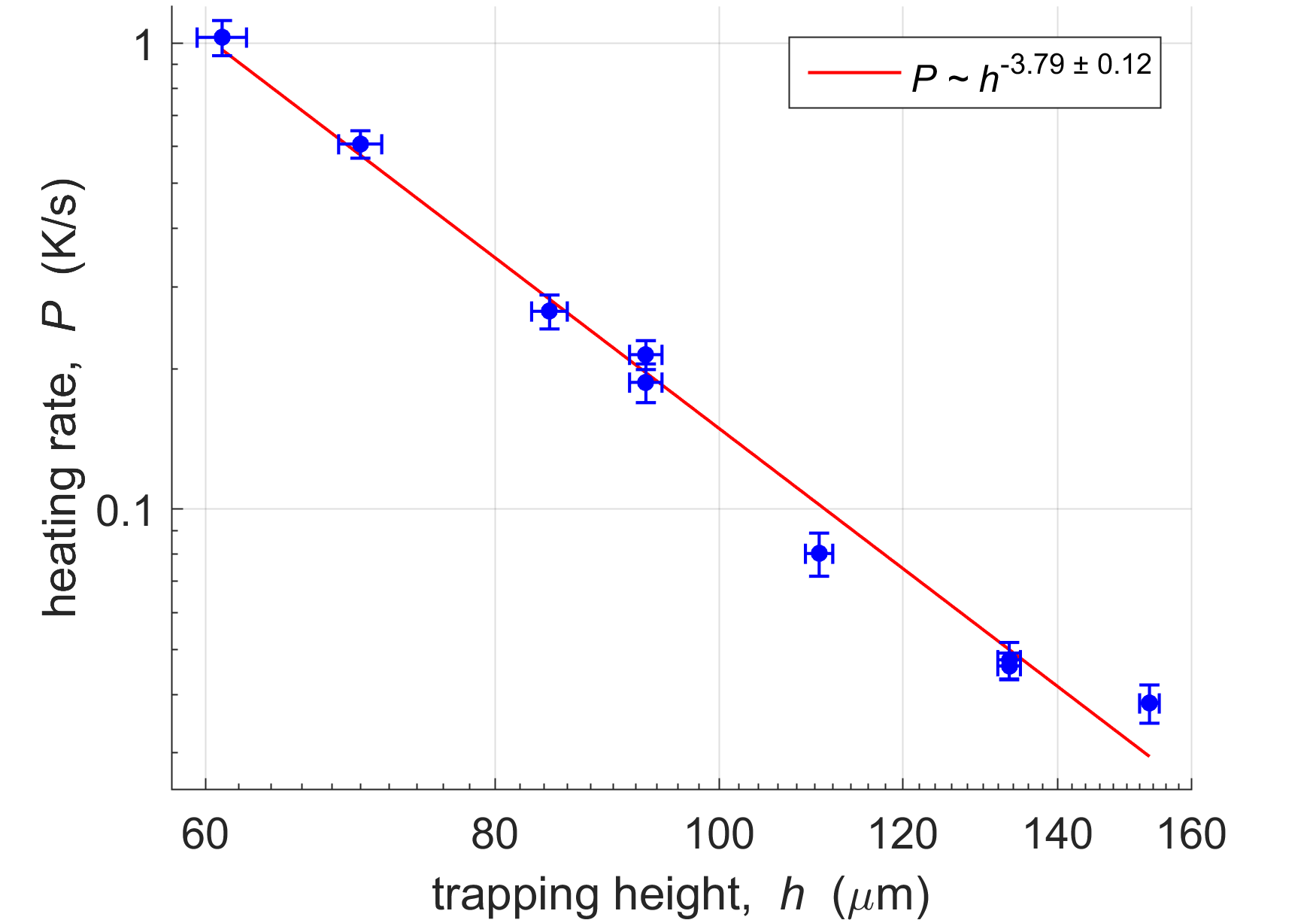}
		\caption{\label{fig:png5} Heating rate (in Kelvin per second) dependence on the trapping height. The axial frequency is 196$\pm$2 kHz for all data points.}
	\end{figure}
	
The heating rate measurement at 154 $\mu$m was done without the voltage divider, the central electrode being grounded directly. A possible reason for the datum for this trapping height in Fig. \ref{fig:png5} being above the fitted line is that other sources of noise, such as externally induced voltage fluctuations, start to play a role at low heating rates. With this point being excluded, the linear fit would give $P \propto h^{-3.99 \pm 0.14}$.
	
	When measuring ion heating rates it is often hard to exclude such factors such as electromagnetic pickup of external fields by loops in the electrodes' circuit or by direct exposure of the ion to external fields \cite{brownnutt2015ion}. Invoking such effects to explain the exponent of -3.79$\pm$0.12 describing the dependence of the heating rate on the trapping height obtained in the experiments reported here seems difficult. We have carried out electric field simulations showing that fluctuations of a potential difference between any two electrodes of our trap would yield electric field fluctuations that would even grow with the trapping height in the range of heights that was used for the measurements (See Supplemental Material \cite{sm}). Therefore, we conclude that the dominant component of the heating rate in our experiments is related to microscopic-scale voltage fluctuations on the electrodes' surfaces and not to external factors such as technical noise.
	
	The power law of the heating rate vs. trapping height $h$ dependence with the power of -3.79$\pm$0.12 that was measured in our experiments is in reasonable agreement with the power of -4 that is often cited \cite{brownnutt2015ion}, though has not been directly measured before. This power law is consistent with the patch potential model \cite{brownnutt2015ion, turchette2000heating, low2011finite} in the limit of small patches. The frequency dependence of the heating rate can be different depending on the mechanism behind the patch potential fluctuations.  This  dependence  has been measured in a large number of experiments and the measured exponents of the power law span from -6 to 1, though most of them concentrate around -1 \cite{brownnutt2015ion}. The exponent measured in our experiments is -1.13$\pm$0.10 which is also close to this value. Another model that fits our experimental results well is the model of a thin dielectric layer covering the electrodes \cite{kumph2016electric}. It predicts the power of -4 for the trapping height dependence of the heating rate and -1 power for the frequency dependence. In conclusion, the assumption of the fluctuating electric field spectral density being proportional to the power of -4 of the distance to the electrode -- important for ion trapping experiments and other areas of experimental physics -- has been for the first time experimentally supported with a direct measurement.
	
	\begin{acknowledgments}
The authors acknowledge funding from Bundesministerium f{\"u}r Bildung und
Forschung (FK 01BQ1012) and from  the European Community's
Seventh Framework Programme under Grant Agreement No. 270843 (iQIT).
	\end{acknowledgments}


	%

	\providecommand{\noopsort}[1]{}\providecommand{\singleletter}[1]{#1}%
	%
	
\pagebreak
\begin{center}
	\textbf{\large Supplemental Material}
\end{center}
\setcounter{equation}{0}
\setcounter{figure}{0}
\setcounter{table}{0}
\setcounter{page}{1}

	
	
	\maketitle
	
	
\begin{figure}[b]
	\includegraphics[width=6cm]{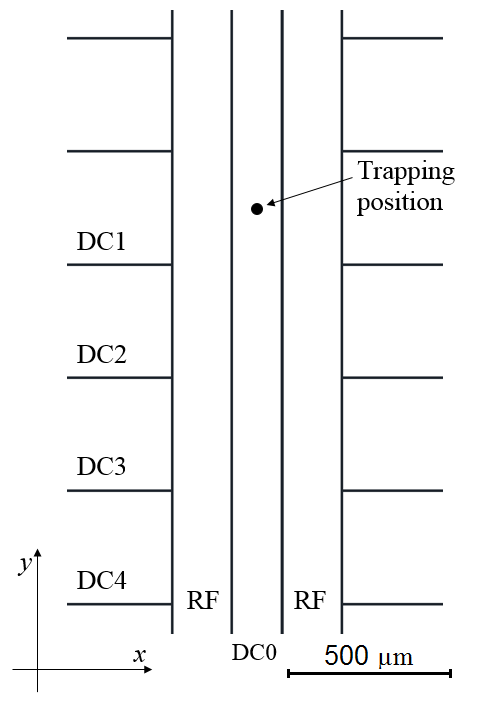}
	\caption{\label{fig:png6} Electrode layout of the surface trap used for the heating rate measurement.}
\end{figure}

\begin{figure}[b]	
	\includegraphics[width=\linewidth]{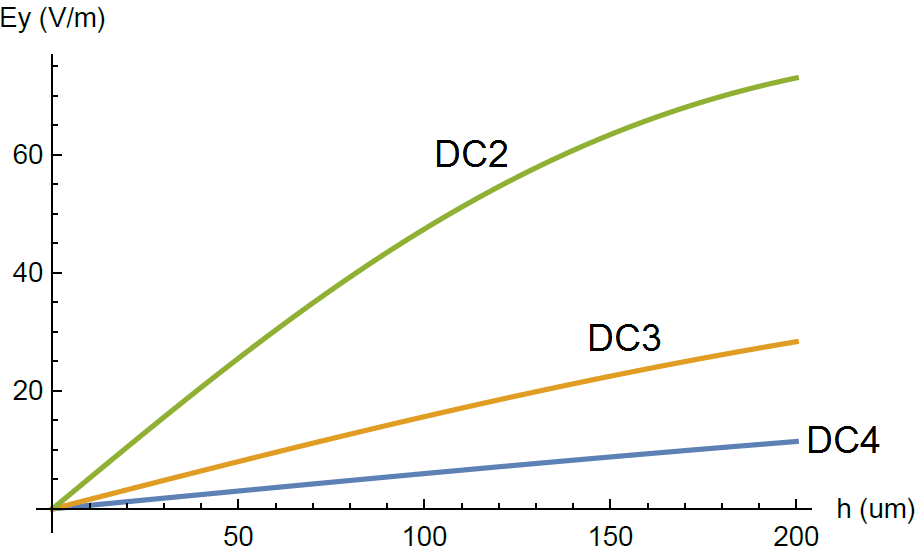}
	\caption{\label{fig:png7} Axial component of the electric field at the trapping position, $E_y$, created by 1 V voltage applied to electrodes DC2, DC3 and DC4 vs. trapping height $h$. DC1, RF and DC0 create electric fields with zero y-component.}
\end{figure}
	
	The sources of electric field fluctuations leading to the heating of ion motion can be divided into two categories:  1) internal, that is their effect cannot be reduced by shielding the setup or filtering the input voltages or somehow improving the devices that provide voltages to the electrodes of the trap; 2) external - all other sources. It is desirable to show that external noise does not significantly contribute to the heating rate in our experiment. For this purpose, we find how external noise would depend on the trapping height $h$. 
	
	Out of the external noise sources, one expects the following ones to be the most significant \cite{brownnutt2015ion}: a) Electromagnetic (EM) interference, that is, excitation of ion motion by any outside EM-fields that penetrate the vacuum chamber; b) EM pickup, that is, voltage fluctuations induced by fluctuations of EM-fields in the loops that may exist between the electrodes of the trap; c) technical noise created by devices that provide voltages to the trap; d) Johnson noise due to any resistances in the voltage supply circuitry. EM interference is not expected to have strong dependence on the trapping height $h$ \cite{brownnutt2015ion}. For the other three noise sources the noise field and hence the ion heating rate dependence on $h$ can be found as follows. 
	
	At the frequency of interest ($\sim$ 1 MHz) the wavelength ($\sim 10^2$ m) is much larger than the electrode dimensions ($\sim 10^{-2}$ m), therefore the electric field created by fluctuations of electrode voltages can be calculated considering the whole electrode to be equipotential and solving the Laplace equation. Therefore the heating rate dependence on the trapping height $h$ will be the same for all of these sources of noise. In order to find it, let $Ey$ be the y-component of the electric field at the trapping position created by applying a constant voltage of 1 V to one of the electrodes and zero voltage to the rest of electrodes. $S_V$ is the spectral density of voltage noise at this electrode. Then the spectral density of the electric field noise $S_E$ can be found as $S_E = S_V \cdot E_y^2$. Therefore, one just needs to find the $E_y(h)$ dependence to obtain the ion axial heating rate due to external noise vs. the trapping height.
	
	The electric field created by a single electrode of our surface trap can be found with an analytical approximation, assuming that the electrode is an equipotential rectangle surrounded by an infinite grounded plane \cite{house2008analytic}. This assumption appears reasonable as the 10 $\mu$m inter-electrode gaps, are much smaller than gap-to-ion distance (95 $\mu$m at minimum) and the trap chip size of 11 mm is much larger than the ion-surface separation that is 154 $\mu$m at maximum. We also compared this analytical model to a numerical simulation of the electric field using SIMION 7.0 software \cite{dahl2000simion} and the results for the electric potential distribution agree within 2\%.

	The electrode layout of our surface trap is shown in Fig.~\ref{fig:png6}. The results of analytical calculations of the axial components of electric field, $E_y$, for a few electrodes of the trap that are closest to the ion are presented in Fig.~\ref{fig:png7}. The DC0, RF and DC1 electrodes create electric fields with zero y-component due to symmetry. In the region of $h$ that was used for the heating rate measurement, $E_y$ grows nearly linearly with the trapping height $h$ for all of the electrodes. That would lead to a quadratic dependence on the trapping height for the ion heating rate which is in strong disagreement with the experimental results. Thus we conclude that external noise is not a dominant factor determining the ion's axial heating rate.

	\providecommand{\noopsort}[1]{}\providecommand{\singleletter}[1]{#1}%

\end{document}